# Quadratic Formulation of Mutual Information for Sensor Placement Optimization using Ising and Quantum Annealing Machines

Yuta Nakano, Shigeyasu Uno
Department of Electrical Systems, Graduate School of Science and Engineering
Ritsumeikan University
E-mail: suno@fc.ritsumei.ac.jp

## Abstract

We address a combinatorial optimization problem to determine the placement of a predefined number of sensors from multiple candidate positions, aiming to maximize information acquisition with the minimum number of sensors. Assuming that the data from predefined candidates of sensor placements follow a multivariate normal distribution, we defined mutual information (MI) between the data from selected sensor positions and the data from the others as an objective function, and formulated it in a Quadratic Unconstrainted Binary Optimization (QUBO) problem by using a method we proposed. As an example, we calculated optimal solutions of the objective functions for 3 candidates of sensor placements using a quantum annealing machine, and confirmed that the results obtained were reasonable. The formulation method we proposed can be applied to any number of sensors, and it is expected that the advantage of quantum annealing emerges as the number of sensors increases.

## 1. Introduction

Combinatorial optimization problems appear in various applications such as shipping, healthcare, and finance [1]. However, these problems are often too complex to be solved accurately in realistic processing time because they are NP-hard [2]. For this reason, there is a strong demand for methods having greater flexibility to take advantage of specific properties of the search space [3].

Metaheuristic algorithms have been known as the methods to solve such complex combinatorial optimization problems within reasonable time limits [4]. With a balance between exploration and exploitation of the search space, such algorithms are effective in solving NP-hard problems [4]. Searching for optimal solutions continues until the solution meets some predefined criterion and a final solution when a system has reached an optimal solution [5].

In metaheuristic algorithms, Ising machine is a statistical-mechanical model of magnetic materials known as the Ising model. This optimizes an objective function with a mathematical formulation known as a Quadratic Unconstrained Binary Optimization (QUBO). In sensor placement problem, the objective is to minimize data reconstruction error and maximize acquired information while minimizing sensor deployment costs [6]. For example, the placement of measurement sensors for temperature in a room can be optimized by modeling with a Gaussian process from a number of candidates [7].

In this work, we propose a new method to derive polynomial objective function for sensor placement optimization problem based on mutual information maximization.

## 2. Mathematical model for multivariate normal distribution

We assume that there are $n$-candidates of sensor placements, and the whole set is denoted as $X$. A random variable $x_i$ denotes observed value at each location of candidates, where $i = 1,2,\cdots,n$. We introduce a column vector $\boldsymbol{x}_X$ consisting of entire set of $x_i$ as follows:

$$\boldsymbol{x}_X = \begin{bmatrix} x_1 \\ x_2 \\ x_3 \\ \vdots \\ x_n \end{bmatrix}. \tag{1}$$

The measured values at each location are assumed to follow normal distributions, and the probability density function is given by

$$P(X) = \frac{1}{\sqrt{(2\pi)^n \det(\boldsymbol{\Sigma}_{XX})}} \exp\left[-\frac{1}{2}(\boldsymbol{x}_X - \boldsymbol{\mu}_X)^\intercal\, \boldsymbol{\Sigma}_{XX}{}^{-1}(\boldsymbol{x}_X - \boldsymbol{\mu}_X)\right], \tag{2}$$

where $\boldsymbol{\mu}_X$ is the column vector of mean $\mu_i$ as follows:

$$\boldsymbol{\mu}_X = \begin{bmatrix} \mu_1 \\ \mu_2 \\ \mu_3 \\ \vdots \\ \mu_n \end{bmatrix}, \tag{3}$$

and $\boldsymbol{\Sigma}_{XX}$ is the covariance matrix as follows:

$$\boldsymbol{\Sigma}_{XX} = \begin{bmatrix} \Sigma_{11} & \Sigma_{12} & \cdots & \Sigma_{1n} \\ \Sigma_{21} & \Sigma_{22} & \cdots & \Sigma_{2n} \\ \vdots & \vdots & \ddots & \vdots \\ \Sigma_{n1} & \Sigma_{n2} & \cdots & \Sigma_{nn} \end{bmatrix}. \tag{4}$$

In Eq.2, $\det(\boldsymbol{\Sigma}_{XX})$ indicates the determinant of $\boldsymbol{\Sigma}_{XX}$.

## 3. Objective function

Here, we denote a set of sensor placement candidates to be selected as $S$, and a set not to be selected as $T$. Then $S$ is the subset of $X$ and $T$ is the difference set. Generally, the entropy of a multivariate normal distribution is given by

$$H(X) = \frac{1}{2}\log[\det(\boldsymbol{\Sigma}_{XX})] + \frac{n}{2}(1 + \log 2\pi). \tag{5}$$

We assume that the entropy of $S$ and $T$ can also be expressed using the same equation, and we define the number of candidates as $n = n_X = n_S + n_T$. Now, mutual information (MI) is the degree of dependence between $S$ and $T$, and is given by

$$I(S,T) = H(S) + H(T) - H(X) = \frac{1}{2}\log\frac{\det(\boldsymbol{\Sigma}_{SS})\det(\boldsymbol{\Sigma}_{TT})}{\det(\boldsymbol{\Sigma}_{XX})}, \tag{6}$$

where $\boldsymbol{\Sigma}_{SS}$ and $\boldsymbol{\Sigma}_{TT}$ are covariance matrices in $S$ and $T$, respectively. In sensor placement problem, we aim to find $S$ which maximizes the MI. We then assign binary bits to express whether an element belongs to $S$ or not. In order to solve this combinatorial optimization problem with the Ising machine, it is necessary to express Eq.6 by such binary variables. As the $\det(\boldsymbol{\Sigma}_{XX})$ is constant regardless of $S$, $\det(\boldsymbol{\Sigma}_{SS})\det(\boldsymbol{\Sigma}_{TT})$ is defined as the objective function $f(S,T)$.

In this paper, we propose an original method to express $f(S,T)$ as QUBO model. First, we define a masking matrix $\boldsymbol{K}$ with $k_{ij}$ $(i,j = 1,2,\dots,n)$ as an element, which takes values as follows:

$$k_{ij}=\begin{cases}1 & i,j\in S\\ 1 & i,j\in T\\ 0 & \text{Otherwise}\end{cases}. \tag{7}$$

Second, we define an element $a_{ij}$ of a matrix $\boldsymbol{A}$ created by overlying (masking) $\boldsymbol{K}$ on $\boldsymbol{\Sigma}_{XX}$ as follows:

$$a_{ij}=\Sigma_{ij}k_{ij}. \tag{8}$$

The rows and columns of $\boldsymbol{A}$ can be swapped to arrive at a matrix $\boldsymbol{B}$ as follows:

$$\boldsymbol{B}=\begin{pmatrix}\boldsymbol{\Sigma}_{SS} & \boldsymbol{0}\\ \boldsymbol{0} & \boldsymbol{\Sigma}_{TT}\end{pmatrix}. \tag{9}$$

Because each row and column are swapped by the same number of times, the total number of swapping will always be even. Therefore, the determinant value remains unchanged and the following equation holds:

$$\det(\boldsymbol{A})=\det(\boldsymbol{B})=\det(\boldsymbol{\Sigma}_{SS})\det(\boldsymbol{\Sigma}_{TT}). \tag{10}$$

By using Leibniz formula, Eq.10 is expressed as follows:

$$\det(\boldsymbol{A})=\sum_{i_1,i_2,\cdots,i_n=1}^{n}\varepsilon_{i_1,i_2,\cdots,i_n}a_{1,i_1}a_{2,i_2}\cdots a_{n,i_n}, \tag{11}$$

where $i_1,i_2,\cdots,i_n$ are assigned so that the values $1,2,\cdots,n$ do no overlap. In addition, using an Ising variable $s_i$ indicating which sensor placement candidate is selected, $k_{ij}$ is expressed as follows:

$$s_i=\begin{cases}+1 & i\in S\\ -1 & i\in T\end{cases}, \tag{12}$$

$$k_{ij}=\frac{1}{2}\left(s_is_j+1\right). \tag{13}$$

Finally, the objective function $f(S,T)$ as a function of $s_i$ is given by

$$\begin{aligned}f(S,T)&=f(s_1,s_2,\cdots,s_n)\\&=\sum_{i_1,i_2,\cdots,i_n=1}^{n}\varepsilon_{i_1,i_2,\cdots,i_n}\Sigma_{1,i_1}\Sigma_{2,i_2}\cdots\Sigma_{n,i_n}k_{1,i_1}k_{2,i_2}\cdots k_{n,i_n}\\&=\sum_{i_1,i_2,\cdots,i_n=1}^{n}\varepsilon_{i_1,i_2,\cdots,i_n}\Sigma_{1,i_1}\Sigma_{2,i_2}\cdots\Sigma_{n,i_n}\frac{1}{2^n}\left(s_1s_{i_1}+1\right)\left(s_2s_{i_2}+1\right)\cdots\left(s_ns_{i_n}+1\right).\end{aligned} \tag{14}$$

Figure 1 shows an example of selecting $S=\{S_2,S_4\}$ when a set of four sensors $X=\{S_1,S_2,S_3,S_4\}$ is given. Figure 1 (a) shows a masking matrix $\boldsymbol{K}$ with $n=4$, as an example, and Fig. 1 (b) shows the covariance matrix $\boldsymbol{\Sigma}_{XX}$ with $n=4$. Black, white, and gray elements represent variance (and covariance) values within $S$, $T$, and the others, respectively. Figure 1 (c) shows the matrix $\boldsymbol{A}$ which is created by overlying (masking) $\boldsymbol{K}$ on $\boldsymbol{\Sigma}_{XX}$, and black and white elements are overlaid 1, and gray elements is overlaid 0, respectively. Figure 1 (d) shows how rows and columns of $\boldsymbol{A}$ are swapped to create matrix $\boldsymbol{B}$**.** Note that, the upper left submatrix of $\boldsymbol{B}$ becomes the covariance matrix within $\boldsymbol{S}$, $\boldsymbol{\Sigma}_{SS}$, and the lower right sub matrix of $\boldsymbol{B}$ becomes that within $\boldsymbol{T}$, $\boldsymbol{\Sigma}_{TT}$, respectively. Thanks to the values of the masking matrix $\boldsymbol{K}$, the upper right and the lower left submatrices are all zero. Therefore, determinant of $\boldsymbol{A}$ is identical to $\det(\boldsymbol{\Sigma}_{SS})\det(\boldsymbol{\Sigma}_{TT})$.

(a)

$$\boldsymbol{K} = \begin{matrix} k11 & k12 & k13 & k14 \\ k21 & k22 & k23 & k24 \\ k31 & k32 & k33 & k34 \\ k41 & k42 & k43 & k44 \end{matrix} = \begin{matrix} 1 & 0 & 1 & 0 \\ 0 & 1 & 0 & 1 \\ 1 & 0 & 1 & 0 \\ 0 & 1 & 0 & 1 \end{matrix}$$

(b)

$$\boldsymbol{\Sigma}_{XX} = \begin{matrix} \Sigma_{11} & \Sigma_{12} & \Sigma_{13} & \Sigma_{14} \\ \Sigma_{21} & \Sigma_{22} & \Sigma_{23} & \Sigma_{24} \\ \Sigma_{31} & \Sigma_{32} & \Sigma_{33} & \Sigma_{34} \\ \Sigma_{41} & \Sigma_{42} & \Sigma_{43} & \Sigma_{44} \end{matrix}$$

(c)

$$\boldsymbol{A} = \begin{matrix} \Sigma_{11} & \Sigma_{12} & \Sigma_{13} & \Sigma_{14} \\ \Sigma_{21} & \Sigma_{22} & \Sigma_{23} & \Sigma_{24} \\ \Sigma_{31} & \Sigma_{32} & \Sigma_{33} & \Sigma_{34} \\ \Sigma_{41} & \Sigma_{42} & \Sigma_{43} & \Sigma_{44} \end{matrix}$$

(d)

$$\boldsymbol{A} = \begin{matrix} \Sigma_{11} & 0 & \Sigma_{13} & 0 \\ 0 & \Sigma_{22} & 0 & \Sigma_{24} \\ \Sigma_{31} & 0 & \Sigma_{33} & 0 \\ 0 & \Sigma_{42} & 0 & \Sigma_{44} \end{matrix} \xrightarrow{\text{Swap rows}} \begin{matrix} 0 & \Sigma_{22} & 0 & \Sigma_{24} \\ 0 & \Sigma_{42} & 0 & \Sigma_{44} \\ \Sigma_{11} & 0 & \Sigma_{13} & 0 \\ \Sigma_{31} & 0 & \Sigma_{33} & 0 \end{matrix}$$

$$\xrightarrow{\text{Swap columns}} \begin{matrix} \boldsymbol{\Sigma}_{SS} \\ \\ \boldsymbol{\Sigma}_{TS} \end{matrix} \begin{matrix} \Sigma_{22} & \Sigma_{24} & 0 & 0 \\ \Sigma_{42} & \Sigma_{44} & 0 & 0 \\ 0 & 0 & \Sigma_{11} & \Sigma_{13} \\ 0 & 0 & \Sigma_{31} & \Sigma_{33} \end{matrix} \begin{matrix} \boldsymbol{\Sigma}_{ST} \\ \\ \boldsymbol{\Sigma}_{TT} \end{matrix} = \boldsymbol{B}$$

Figure. 1. The schematic diagram swapping rows and columns from matrices $\boldsymbol{K}$, $\boldsymbol{\Sigma}_{XX}$ to $\boldsymbol{B}$. (a) Masking matrix $\boldsymbol{K}$ with $n = 4$. (b) The covariance matrix $\boldsymbol{\Sigma}_{XX}$ with $n = 4$. (c) The composition overlying (masking) $\boldsymbol{K}$ on $\boldsymbol{\Sigma}_{XX}$. (d) The flow that $\boldsymbol{A}$ transforms into $\boldsymbol{B}$.

## 4. Validation and result using toy models

In this section, we validate the formula presented in Sec.3 using toy models with 3 sensors. The objective function $f(s_1, s_2, s_3)$ in this case is as follows:

$$f(s_1, s_2, s_3) = \frac{1}{2^3} \sum_{i_1, i_2, i_3 = 1}^{3} \varepsilon_{i_1, i_2, i_3} \Sigma_{1, i_1} \Sigma_{2, i_2} \Sigma_{3, i_3} (s_1 s_{i_1} + 1)(s_2 s_{i_2} + 1)(s_3 s_{i_3} + 1). \tag{15}$$

Although 4th-order and 6th-order terms of Ising variables appear, they can be reduced to 2nd-order or lower for the following reasons. The 6th-order term becomes constant because $i_1, i_2, i_3$ are assigned to $1, 2, 3$, and it leads to $s_1 s_{i_1} s_2 s_{i_2} s_3 s_{i_3} = {s_1}^2 {s_2}^2 {s_3}^2 = 1$. The 4th-order terms such as $s_1 s_{i_1} s_2 s_{i_2} + s_2 s_{i_2} s_3 s_{i_3} + s_3 s_{i_3} s_1 s_{i_1}$ can be reduced to 2nd-order or lower because the same Ising variable appears at least once in each term when $i_1, i_2, i_3$ are assigned to $1, 2, 3$. For instance, $(i_1, i_2, i_3) = (2, 1, 3)$, the 4th-order term is given by

$$s_1 s_{i_1} s_2 s_{i_2} + s_2 s_{i_2} s_3 s_{i_3} + s_3 s_{i_3} s_1 s_{i_1} = 1 + 2 s_1 s_2. \tag{16}$$

Based on these reductions, Eq.15 can be simplified down to 2nd-order terms:

$$\begin{aligned} f(s_1, s_2, s_3) = \Sigma_{11}\Sigma_{22}\Sigma_{33} + \frac{\Sigma_{12}\Sigma_{23}\Sigma_{31}}{2}(1 + s_1 s_2 + s_2 s_3 + s_3 s_1) \\ -\Sigma_{12}\Sigma_{33}(1 + s_1 s_2) - \Sigma_{23}\Sigma_{11}(1 + s_2 s_3) - \Sigma_{31}\Sigma_{22}(1 + s_3 s_1). \end{aligned} \tag{17}$$

Thus the problem is reduced to a QUBO problem, and it is possible to solve the problem using Ising machine.

When implementing the problem in a Ising machine, we minimize $-f(s_1, s_2, s_3)$ to search for the ground-state energy. Ising machines used in this work are Fixstars Amplify AE [8] and D-Wave Quantum Annealer [8], and the execution time and the number of iterations are specified as 1000ms and 100, respectively.

## 4.1 Toy model A

As an example where a strong correlation is created between the sensors $S_1$ and $S_3$, and the sensor $S_2$ has a weak correlation with the others, consider the covariance matrix given by

$$\boldsymbol{\Sigma}_{XX} = \begin{bmatrix} 2 & 0.1 & 1 \\ 0.1 & 2 & 0.1 \\ 1 & 0.1 & 2 \end{bmatrix}. \tag{18}$$

Obviously, it is optimal to select $S_1, S_2$ or $S_2, S_3$, and theoretical output value of $-f(s_1, s_2, s_3)$ is -7.98. In Figure 2, the strength of the correlations between sensors is represented by the thickness of the double-headed arrows.

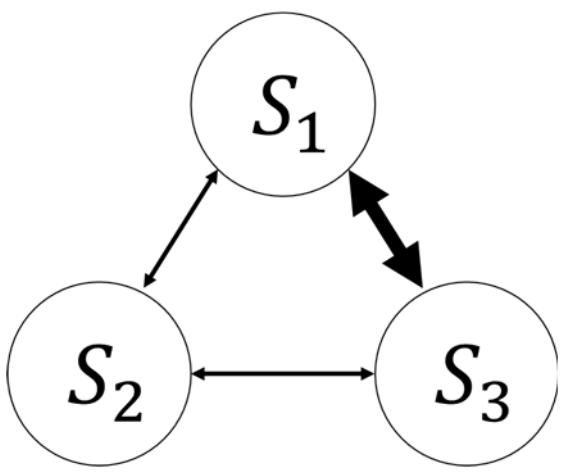

Figure. 2. Schematic diagram of data correlation for the toy model A.

Table. 1. Calculation result by Fixstars amplify AE.

| $[s_1, s_2, s_3]$ | Output value |
|---|---|
| $[1, 1, -1]$ | $-7.60$ |
| $[-1, 1, 1]$ | $-7.60$ |
| $[-1, -1, 1]$ | $-7.60$ |
| $[1, -1, -1]$ | $-7.60$ |

Table. 2. Calculation result by D-Wave Quantum Annealer.

| $[s_1, s_2, s_3]$ | Output value | Output count |
|---|---|---|
| $[1, 1, -1]$ | $-7.60$ | 24 |
| $[-1, 1, 1]$ | $-7.60$ | 14 |
| $[-1, -1, 1]$ | $-7.60$ | 29 |
| $[1, -1, -1]$ | $-7.60$ | 33 |

Tables 1 and 2 show optimization results by Fixstars amplify AE and D-Wave Quantum Annealer, respectively. Note that the solutions $[s_1, s_2, s_3] = [1, -1, -1]$ and $[-1, -1, 1]$are equivalent to $[s_1, s_2, s_3] = [-1, 1, 1]$ and $[1, 1, -1]$ respectively because the Ising variables are grouped based on whether they take the value or 1 of -1 in the calculation of $-f(s_1, s_2, s_3)$. Thus, all the solutions from Ising and quantum annealing machines successfully indicate all the possible optimal solutions.

## 4.2 Toy model B

When strong correlations are created in the following order: between $S_1$ and $S_3$, between $S_1$ and $S_2$, and between $S_2$ and $S_3$, as depicted in Fig.3, the covariance matrix is given by

$$\boldsymbol{\Sigma}_{XX} = \begin{bmatrix} 2 & 0.5 & 1 \\ 0.5 & 2 & 0.1 \\ 1 & 0.1 & 2 \end{bmatrix} \tag{19}$$

Similar to the toy model A, it is optimal to select $S_2$ or $S_3$, and the output value of $-f(s_1, s_2, s_3)$ is -7.98.

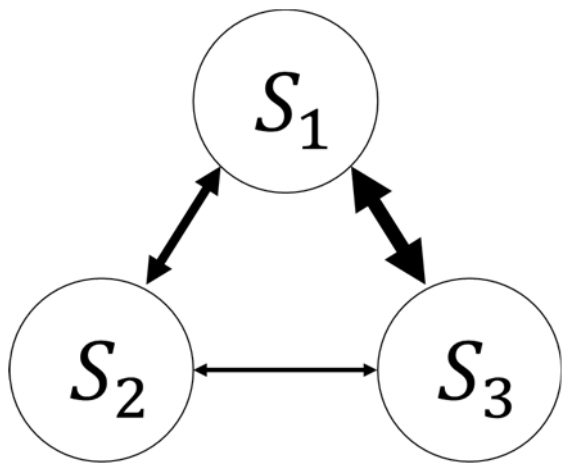

Figure. 3. Schematic diagram of data correlation for the toy model B.

Table. 3. Calculation result by Fixstars amplify AE.

| $[s_1, s_2, s_3]$ | Output value |
|---|---|
| $[-1, 1, 1]$ | $-7.60$ |
| $[1, -1, -1]$ | $-7.60$ |

Table. 4. Calculation result by D-Wave Quantum Annealer.

| $[s_1, s_2, s_3]$ | Output value | Output count |
|---|---|---|
| $[-1, 1, 1]$ | $-7.60$ | 43 |
| $[1, -1, -1]$ | $-7.60$ | 54 |
| $[1, 1, -1]$ | $-6.0$ | 2 |
| $[-1, 1, -1]$ | $-4.0$ | 1 |

Tables 3 and 4 show optimization results by Fixstars amplify AE and D-Wave Quantum Annealer, respectively. In most cases, the optimal combinations according to theory are resulted, however the results by D-Wave Quantum Annealer output combinations which are not optimal, such as $[s_1, s_2, s_3] = [1,\ 1,\ -1], [-1,\ 1,\ -1]$, which may be due to errors occurred during the calculation of the optimal solution search. The output values of non-optimal solutions are higher than ones of optimal solutions because they did not reach the ground state.

Thus, in a scale with three sensors, such as the toy models A and B, we confirm that MI can be maximized and the optimal combinations of sensors can be outputted by Fixstars amplify AE and D-Wave Quantum Annealer. Thus, it has been demonstrated that the optimization calculations in the Ising model based on our proposed formula provide correct solutions.

## 5. Discussion

In this work, we conducted sensor placement optimization using our proposed formula based on MI. We have proposed a method of formulating it to QUBO. Meanwhile, the optimization often becomes Higher Order Binary Optimization (HOBO) involving functions with higher-order terms when the number of sensors is four or more. In the case of four sensors, for example, terms of 8th, 6th, and 4th orders will appear. One approach to reduce such HOBO to QUBO is to determine the number of selected sensors in advance and add associated constraint terms to the objective function. However, manual conversion of high-order terms into quadratic form becomes impractical as the number of sensors increases. To avoid this, methods are needed to either split the variables most frequently occurring in higher-order terms and reduce them to quadratic form or to add auxiliary variables to convert higher-order terms into equivalent quadratic terms [9]. It is necessary to propose a suitable method depending on the scale of the problem.

Our future goal is to find out how to generalize functions containing higher-order terms into quadratic terms. Originally, the sensor placement optimization problem aims to reduction of the number of sensors while maximizing measurement performance, seeking Pareto optimal solutions. In the current scale of three sensors, there is no need to distinguish the number of selected sensors, as their MI calculations are identical. For cases with four or more sensors, as mentioned earlier, one approach is to pre-determine the number of sensors to be selected and then output the optimal solution among all patterns, aiming for the minimum values. Furthermore, it becomes possible to achieve Pareto optimization in a single operation by adding constraint terms to the objective function that favors better values as fewer sensors are selected. In other words, the importance of considering constraints that limit the number of selected sensors becomes higher with increasing number of sensors.

The method proposed in this work is intended for application to sensor placement optimization problems. However, it can also be applied to the other mathematical models involving the selection of elements from systems represented by the multivariate normal distributions with correlations. Moreover, masking matrix method can be applied if the objective function involves calculations with determinants.

## 6. Conclusions

In this work, we proposed a new approach to formulate the problem of finding the optimal sensor placement from candidates represented by a multivariate normal distribution with correlations into a quadratic form, and validate its effectiveness using quantum annealing. The verification results using toy models confirmed that the formulation method proposed in this work is applicable for utilizing quantum annealing in sensor placement optimization problems. The challenge ahead is to establish an effective method for transforming from HOBO to QUBO, and enable the derivation of Pareto optimal solutions when the number of sensors increases.